\documentclass{llncs}

\usepackage[utf8]{inputenc}
\usepackage[T1]{fontenc}
\usepackage[brazilian]{babel}
\usepackage{csquotes}
\usepackage{hyphenat}
\usepackage{bm}
\usepackage{amssymb}
\usepackage{amsmath}
\usepackage{array}
\usepackage{mathrsfs}
\usepackage{booktabs}
\usepackage{appendix}

\usepackage[locale=FR]{siunitx}

\usepackage{graphicx}
\graphicspath{ {images/} }

\usepackage{listings}
\newcommand{\upcite}[1]{\textsuperscript{\textsuperscript{\cite{#1}}}}

\begin{document}

\title{Network Vector Autoregressive Model for Dyadic Response Variables}
%

\author{Jiajia Wang}
%
%
%
\institute{School of Mathematics , Yunnan Normal University\\
\email{tomatowjj@126.com}}

\maketitle              
\renewcommand\refname{References}
\renewcommand{\abstractname}{Abstract}
\begin{abstract}
For general panel data, by introducing network structure, network vector autoregressive (NVAR) model captured the linear inter dependencies among multiple time series. In this paper, we propose network vector autoregressive model for dyadic response variables (NVARD), which describes the dynamic process of dyadic data in the case of the dependencies among different pairs are taken into consideration. Besides, due to the existence of heterogeneity between time and individual, we propose time-varying coefficient network vector autoregressive model for dyadic response variables (VCNVARD). Finally, we apply these models to predict world bilateral trade flows.
\end{abstract}

\section{Introduction}
Dyadic data consist of measurements made on pairs of objects. So that $y_{ij}$ denotes the value of measurement on the potentially ordered pair $(i,j)$. The measurements can be direct or undirect. Dyadic data are also called network data, that exist in many areas, such as social network\upcite{Zhu2017}, protein interaction network\upcite{Raftery2012}, world trade network\upcite{Ward2013} and so on. For example, in a direct social network, $y_{ij}=1$ represents that $i$ claims $j$ is her or his friend or neighbour, $y_{ij}=0$ represents the opposite. In a dynamic world trade network, $y_{ijt}$ represents the trade volume that country $i$ export to country $j$ at time $t$.

The existing literatures are relatively perfect about the statistical model of non dyadic response variables\upcite{Wooldridge2010,Hsiao2003}. But for dyadic response variables, there are still many questions need to be discussed.

Let $y_{it}$ be the continuous response obtained from node $i$ at time $t$, then $\mathbf{y}_t=(y_{1t},y_{2t},\cdots,y_{nt})'$ constitutes an high dimensional vector, its time series dynamic needs to be statistically modeled. We can model each individual time series separately, but the relationship across different time series is lost. We can also model it by a vector autoregressive model (VAR)\upcite{Hamilton1994}. However, the number of parameters need to be estimated may be too large. Zhu et.al. (2017)\upcite{Zhu2017} introduced a social network into model. In their views, $y_{it}$ might be affected by its neighbor in social network. As a result, they proposed a network vector autoregressive (NVAR) model which assumed that $y_{it}$ is a linear combination of the momentum effect, the network effect, the nodal effect and an independent noise. Compared with a usual VAR model, its total number of unknown parameters is fixed.

For dyadic response variables, considering the dependency structure between different cross sections, some articles introduced latent variables\upcite{Hoff2002,Sewell2015,Durante2014,Nowicki2001}. But they still have some limits, such as estimation and computational problems and so on\upcite{Hunter2012}.

\par When carrying on this paper, we have two main objectives: (1) Modelling dynamic dyadic responses; (2) Predicting future relationships between individuals. Based on this, we propose network vector autoregressive model for dynamic dyadic response variables (NVARD). Additionally, we introduce state variables to take into account the heterogeneity of panel data and propose time-varying coefficient network vector autoregressive model for dynamic dyadic response variables (VCNVARD).

The remainder of the article is organized as follows: Section 2 introduces the models that we propose. Section 3 outlines the Bayesian estimation method of the model parameters and state variables. Section 4 describes how to obtain network predictions. Section 5 presents the results from analyzing world trade data. Section 6 provides a brief discussion.

%
\section{The Models\label{s2}}

\par For general dynamic response variable $y_{it}\ (i\in\{1,\cdots,n\},t\in\{1,\cdots,T\})$, taking the social network structure into account, network vector autoregressive model had been proposed\upcite{Zhu2017}
\begin{equation}Y_{it}=\beta_0+Z'_i\gamma+\beta_1\sum_{j\neq i}\frac{a_{ij}}{\sum_{j\neq i}a_{ij}}Y_{j(t-1)}+\beta_2Y_{i(t-1)}+\varepsilon_{it}\label{for1}\end{equation}
where $a_{ij}=1$ if there exists a social relationship from $i$ to $j$ and $a_{ij}=0$ otherwise. Moreover, $Z_i=(Z_{i1},\cdots,Z_{iM})'\in R^M$ is an M dimensional node-specific random vector that can be observed. Besides, they argued that if $E||Z_i||<\infty$ and $|\beta_1|+|\beta_2|<1$, then there existed a unique strictly stationary solution with finite first order moment. Meanwhile , under some regular conditions, the ordinary least squares estimator is consistent and asymptotically normal when $T$ is fixed, $n\rightarrow\infty$ or $min\{T,n\}\rightarrow\infty$.
\par However, there are still some topics need further research. On the one hand, how to characterize the inter dependencies among response variables in a statistical model for dyadic response $y_{ijt}$. On the other hand, when involving both time series and cross sectional data, heterogeneity need to be taken into account.
The article will elaborate the first aspect in Subsection \ref{s2.1}, the second aspect in Subsection \ref{s2.2}.

\subsection{Network Vector Autoregressive Model for Dyadic Response Variables (NVARD)\label{s2.1}}
\par Recall that $n$ is the network size which is the number of nodes and $y_{ijt}\ (i,j\in\{1,\cdots,n\},t\in\{1,\cdots,T\})$ is response variable which represents the directed relationship from node $i$ to $j$ at time $t$. As in common articles, we assume that a node cannot have a relationship with itself, that is $y_{ii}$ is not defined. In addition, for each node $i$, assume that $X_{ijt}=(X_{ijt1},\cdots,X_{ijtM})'\in R^M$ is non-random attributes of node $i$, node $j$ or pair $(i,j)$ that can be obtained at time $t-1$. Our target is to construct a model for $y_{ijt}$.
\par Under network framework, dyadic response $y_{ijt}$ may be affected by three different factors except independent noise. Firstly, $y_{ijt}$ may be affected by itself but from the previous time point, which is $y_{ij(t-1)}$. Secondly, $y_{ijt}$ may be affected by a set of covariates $X_{ijt}$, which are non-random, and can be observed at time $t-1$. They contain attributes of node $i$, attributes of node $j$, attributes of pair $(i,j)$, and all these can be related to time. Thirdly, $y_{ijt}$ may be affected by other pairs $(k,l)$ who own node $i$ or $j$, that are, inverse flow $y_{ji}$, origin-origin flows $y_{i1},\cdots,y_{in}$, origin-destination flows $y_{1i},\cdots,y_{ni}$, destination-origin flows $y_{j1},\cdots,y_{jn}$, destination-destination flows $y_{1j},\cdots,y_{nj}$. So network vector autoregressive model for dyadic response variables (NVARD) has the form
\begin{eqnarray}
y_{ijt}&=&\beta_1+\beta_2y_{ij(t-1)}+\beta_3y_{ji(t-1)}+\beta_4\sum_{k\neq i,j} w_{ijk}^{(oo)}y_{ik(t-1)}+\beta_5\sum_{k\neq i,j}w_{ijk}^{(od)}y_{ki(t-1)}\nonumber\\
&&+\beta_6\sum_{k\neq i,j}w_{ijk}^{(do)}y_{jk(t-1)}+\beta_7\sum_{k\neq i,j}w_{ijk}^{(dd)}y_{kj(t-1)}+X_{ijt}'\beta_8+\varepsilon_{ijt}\label{for2}
\end{eqnarray}
where $\beta_1,\beta_2,\beta_3,\beta_4,\beta_5,\beta_6,\beta_7$ are scalars, $\beta_8$ is a vector, and $\beta_8=(\beta_{81},\cdots,\beta_{8M})'$. From the above formula, we can see that the total number of covariates is $K=7+M$.
\par Though the network is not symmetric, but we believe that the relationship between $y_{ij}$ and $y_{ji}$ is quite closed, so we deal with it separately. $w$ denote the normalized weights, that are $\sum_{k\neq i,j} w_{ijk}^{(oo)}=1$, $\sum_{k\neq i,j}w_{ijk}^{(od)}=1$, $\sum_{k\neq i,j}w_{ijk}^{(do)}=1$, $\sum_{k\neq i,j}w_{ijk}^{(dd)}=1$. Additionally, we assume these weights are exogenous.
\par In order to make the model and estimation more intuitive, we simplify the model, that is, the variables are expressed in the form of vectors or matrices. We introduce $\text{vecd}(A=\{a_{ij}\})$ to denote the vectorization of matrix A excluding diagonal elements, that is,
$$\text{vecd}(A)=(a_{21},a_{31},\cdots,a_{n1},a_{12},\cdots,a_{n2},\cdots,a_{(n-1)n})'.$$
Correspondingly, formula (\ref{for2}) can be rewritten as
\begin{equation}Y_t=Z_t\bm{\beta}+\bm{\varepsilon}_t,\bm{\varepsilon}_t\sim N(0,\sigma^2_{\varepsilon}I)\label{for3}\end{equation}
where $Y_t$, $\bm{\varepsilon}_t$ is vectorization of matrix,
$$Y_t=\text{vecd}\left(\left[
                   \begin{array}{ccc}
                     Y_{11t} & \cdots & Y_{1nt} \\
                     \vdots & \ddots & \vdots \\
                     Y_{n1t} & \cdots & Y_{nnt} \\
                   \end{array}
                 \right]\right)
,\ \bm{\varepsilon}_t=\text{vecd}\left(\left[
                   \begin{array}{ccc}
                     \varepsilon_{11t} & \cdots & \varepsilon_{1nt} \\
                     \vdots & \ddots & \vdots \\
                     \varepsilon_{n1t} & \cdots & \varepsilon_{nnt} \\
                   \end{array}
                 \right]\right).$$
$I$ denotes identity matrix with compatible dimension, $\bm{\beta}=(\beta_1,\beta_2,\cdots,\beta_7,\beta_8')'$ is coefficient vector, covariates $Z_t=\left(Z'_{21t},Z'_{31t},\cdots,Z'_{(n-1)nt}\right)'$ is a $(n^2-n)\times K$ matrix, and
\begin{eqnarray*}
Z_{ijt}&=&\left(y_{ij(t-1)},y_{ji(t-1)},\sum_{k\neq i,j}w_{ijk}^{(oo)}y_{ik(t-1)},\sum_{k\neq i,j}w_{ijk}^{(od)}y_{ki(t-1)},\right.\\
&&\left.\sum_{k\neq i,j}w_{ijk}^{(do)}y_{jk(t-1)},\sum_{k\neq i,j}w^{(dd)}_{ijk}y_{kj(t-1)},X'_{ijt}\right).
\end{eqnarray*}

\subsection{Time-Varying Coefficient Network Vector Autoregressive Model for Dyadic Responses (VCNVARD)\label{s2.2}}

To eliminate the heterogeneity of time series and cross sectional data, the varying intercept model is widely used\upcite{Hsiao2003,Greene2008}. It should be noted that, for dyadic responses, the time span $T$ is usually small. On the contrary, the number of cross sectional individual is usually quite large, that is $O(n^2)$. In this case, adding the individuals effects into a model may lead to over-fitting. So we introduce time-varying effects into the model. Since there are dynamic components, namely, the lagged term, it is considered that some coefficients are time-dependent and other coefficients are unchanged. Therefore, model (\ref{for3}) can be written as
\begin{equation}y_{ijt}=Z^{(1)}_{ijt}\bm{\beta}^{(1)}+Z^{(2)}_{ijt}\bm{\beta}^{(2)}_t+\varepsilon_{ijt},\ t=1,\cdots,T\label{for4}\end{equation}
where $\varepsilon_{ijt}$ is error term $\varepsilon_{ijt}\sim N(0,\sigma^2_{\varepsilon})$, and independent of each other for different pairs $(i,j)$ and time $t$. Superscript $(1)$ describes the covariates whose coefficients are invariant, and superscript $(2)$ is the opposite.
\par Based on the above analysis and the aim of prediction, we use state variables to describe time-varying coefficients, named time-varying coefficient model of network vector autoregression for dyadic response variables (VCNARD),
\begin{equation}\left\{\begin{array}{l}
           {\rm Observation\ Equation}:Y_t=Z^{(1)}_{t}\bm{\beta}^{(1)}+Z^{(2)}_{t}\bm{\beta}^{(2)}_t+\bm{\varepsilon}_{t},\ t=1,\cdots,T \\
           {\rm State\ Equation}:\bm{\beta}^{(2)}_{t}=\bm{\beta}^{(2)}_{t-1}+\mathbf{u}_t,\ t=2,\cdots,T
         \end{array}\right.\label{for5}
\end{equation}
where $Z^{(2)}_{t}$ contains part columns of $Z_t$ whose coefficients are varying over time, $Z^{(1)}_{t}$ contains the rest columns of $Z_t$. $\bm{\beta}^{(2)}$ contains varying coefficients and $\bm{\beta}^{(1)}$ contains the rest coefficients. Besides, $\mathbf{u}_t\sim N(0,\sigma^2_uI)$ are independent of each other for different $t$.
Other symbols are the same as Subsection \ref{s2.1}.

\section{Bayesian Estimation\label{s3}}
\par There are many methods to estimate parameters in NVARD, such as ordinary least squares method, maximum likelihood method. However, since there are some state variables in VCNVAR, Bayesian method can be used to estimate both parameters and state variables. Bayesian action can minimize Bayesian risk. Consequently, if the loss function is the a square error loss function, the Bayesian estimator is posterior expectation.

\subsection{Estimation for NVARD\label{s3.1}}
\par In order to avoid using too many hyper-parameters, we use improper prior distribution.
\par (1) For variance, the improper prior distributions are
$$\sigma^2_{\epsilon}\propto\frac{1}{\sigma^2_{\epsilon}}$$
\par (2) For coefficients of linear model, the improper joint prior distribution is
$$\bm{\beta|}\sigma^2_{\varepsilon}\propto 1$$
\par We assume that $Y_0$ is known, according to the formula (\ref{for3}), the posterior joint distribution of $\bm{\beta}$,$\sigma^2_{\varepsilon}$ is
\begin{eqnarray*}
&&p(\bm{\beta},\sigma^2_{\varepsilon}|Y_1,\cdots,Y_T)\propto p(\bm{\beta},\sigma_\varepsilon^2,Y_1,\cdots,Y_T)\\
&\propto&p(\sigma^2_{\varepsilon})p(\bm{\beta}|\sigma^2_{\varepsilon})\prod^T_{t=1}p(Y_t|Y_1,\cdots,Y_{t-1},\bm{\beta},\sigma^2_{\varepsilon})\\
&\propto&(\sigma^2_{\varepsilon})^{-1}(\sigma^2_{\varepsilon})^{-\frac{(n^2-n)T}{2}}\exp\left\{-\frac{\sum^T_{t=1}(Y_t-Z_t\bm{\beta})'(Y_t-Z_t\bm{\beta})}{2\sigma^2_{\varepsilon}}\right\}
\end{eqnarray*}
Hence, the posterior marginal distribution of $\bm{\beta}$ is
\begin{equation}
p(\bm{\beta}|Y_1,\cdots,Y_T)\sim Mt(v,\mu,\Sigma).\label{for6}
\end{equation}
\par The details of formula (\ref{for6}) are given in Appendix \ref{A}. Where $Mt$ represents multivariate t distribution. The first parameter $v$ is freedom degree of multivariate t distribution, which equals $(n^2-n)T-K$. The second parameter $\mu$ is mean of multivariate t distribution, which equals $\left(\sum^T_{t=1}Z'_tZ_t\right)^{-1}\sum^T_{t=1}Z'_tY_t$. The third parameter $\Sigma$ is precision matrix,
$$\Sigma=\frac{1}{v}\left(\sum^T_{t=1}Z'_tZ_t\right)^{-1}\left(\sum^T_{t=1}Y'_tY_t-(\sum^T_{t=1}Z'_tY_t)'(\sum^T_{t=1}Z'_tZ_t)^{-1}(\sum^T_{t=1}Z'_tY_t)\right).$$
\par The posterior marginal distribution of $\sigma^2_{\varepsilon}$ is
\begin{equation}
p(\sigma^2_{\varepsilon}|Y_1,\cdots,Y_T)\sim IG(a,b)\label{for7}
\end{equation}
where $a$ is the shape parameter, whose value is $a=\frac{(n^2-n)T-K}{2}$, $b$ is the rate parameter, whose value is
$$b=\frac{\sum^T_{t=1}Y'_tY_t-(\sum^T_{t=1}Z'_tY_t)'(\sum^T_{t=1}Z'_tZ_t)^{-1}(\sum^T_{t=1}Z'_tY_t)}{2}.$$
\par The details of formula (\ref{for7}) are given in Appendix \ref{B}.
Therefore, the estimates of parameter $\bm{\beta}$ and $\sigma^2_{\varepsilon}$ are
$$\hat{\bm{\beta}}=E(\bm{\beta}|Y_1,\cdots,Y_T)=\mu=\left(\sum^T_{t=1}Z'_tZ_t\right)^{-1}\sum^T_{t=1}Z'_tY_t,$$
$$\hat{\sigma}^2_{\varepsilon}=E(\sigma^2_{\varepsilon}|Y_1,\cdots,Y_T)=\frac{b}{a-1}=\frac{\sum^T_{t=1}Y'_tY_t-(\sum^T_{t=1}Z'_tY_t)'(\sum^T_{t=1}Z'_tZ_t)^{-1}(\sum^T_{t=1}Z'_tY_t)}{(n^2-n)T-K-2}.$$

\subsection{Estimation for VCNVARD\label{s3.2}}

\par Similar to Subsection \ref{s3.1}, we still use improper prior distribution,
\par (1) For variance, the improper prior distributions is
$$p\left(\sigma^2_{\epsilon},\sigma^2_u\right)\propto\frac{1}{\sigma^2_{\epsilon}}\cdot\frac{1}{\sigma^2_u}$$
\par (2) For constant coefficients of linear model, the improper joint prior distribution is
$$p\left(\bm{\beta}^{(1)}|\sigma^2_{\varepsilon},\sigma^2_u\right)\propto 1$$
\par (3) For the initial position of state variable, the improper prior distribution is
$$p\left(\bm{\beta}^{(2)}_1|\sigma^2_{\varepsilon},\sigma^2_u,\bm{\beta}^{(1)}\right)\propto 1$$
\par According to the formula (\ref{for5}), the posterior joint distribution of parameters and state variables is
\begin{eqnarray*}
&&p(\sigma^2_{\varepsilon},\sigma^2_u,\bm{\beta}^{(1)},\bm{\beta}^{(2)}_1,\cdots,\bm{\beta}^{(2)}_T|Y_1,\cdots,Y_T)\\
&\propto&p\left(\sigma^2_{\varepsilon},\sigma^2_u\right)p\left(\bm{\beta}^{(1)}|\sigma^2_{\varepsilon},\sigma^2_u\right)p\left(\bm{\beta}^{(2)}_1|\sigma^2_{\varepsilon},\sigma^2_u,\bm{\beta}^{(1)}\right)
\prod^T_{t=2}p(\bm{\beta}^{(2)}_t|\bm{\beta}^{(2)}_1,\cdots,\bm{\beta}^{(2)}_{t-1},\sigma^2_{\varepsilon},\sigma^2_u,\bm{\beta}^{(1)})\\
&&\cdot\prod^T_{t=1}p(Y_t|Y_1,\cdots,Y_{t-1},\bm{\beta}^{(2)}_1,\cdots,\bm{\beta}^{(2)}_{T},\sigma^2_{\varepsilon},\sigma^2_u,\bm{\beta}^{(1)})\\
&\propto&\frac{1}{\sigma^2_{\epsilon}}\cdot\frac{1}{\sigma^2_u}\prod^T_{t=2}(\sigma^2_u)^{-\frac{m}{2}}\exp\left\{-\frac{(\bm{\beta}^{(2)}_t-\bm{\beta}^{(2)}_{t-1})'(\bm{\beta}^{(2)}_t-\bm{\beta}^{(2)}_{t-1})}{2\sigma^2_u}\right\}\\
&&\cdot\prod^T_{t=1}\prod_{i\neq j}(\sigma^2_{\varepsilon})^{-\frac{1}{2}}\exp\left\{-\frac{\left(y_{ijt}-Z^{(1)}_{ijt}\bm{\beta}^{(1)}-Z^{(2)}_{ijt}\bm{\beta}^{(2)}_t\right)^2}{2\sigma^2_{\varepsilon}}\right\}
\end{eqnarray*}
where $m$ is the number of varying coefficients.
\par It is difficult to obtain posterior marginal distributions for each variable by integration. However, if we can draw samples from the posterior distribution, the sample mean converges to the posterior mean according to the law of large numbers. That is
 $$\frac{1}{L}\sum^L_{l=1}\Psi^l\ \xrightarrow{\ p\ } E(\Psi|Y_1,\cdots,Y_T)$$
 where $\Psi$ denotes parameters and all state variables, $\Psi^l$ represents the $l$th sample draws from the distribution $p(\Psi|Y_1\cdots,Y_T)$.
 \par Markov chain Monte Carlo (MCMC) methods are a class of algorithms for sampling from a probability distribution based on constructing a Markov chain that has the desired distribution of its equilibrium distribution\upcite{Carter1994,Geweke2001}. The state of the chain after a number of steps is then used as a sample of the desired distribution. Hence, we use Gibbs algorithm to draw samples. Necessarily, we need to compute the full conditional distribution of each part. The formulas and computational methods for the full conditional distributions are given in Appendix \ref{C}. Therefore, the estimates of parameter and state variables are
$$\hat{\bm{\beta}}^{(1)}=E\left[\bm{\beta}^{(1)}|Y_1,Y_2,\cdots,Y_T\right]\approx \frac{1}{L}\sum^L_{l=1}(\bm{\beta}^{(1)})^l$$
$$\hat{\bm{\beta}}^{(2)}_t=E\left[\bm{\beta}^{(2)}_t|Y_1,Y_2,\cdots,Y_T\right]\approx \frac{1}{L}\sum^L_{l=1}(\bm{\beta}_t^{(2)})^l,\ t=1,2,\cdots,T$$
$$\hat{\sigma}^2_{\varepsilon}=E(\sigma^2_{\varepsilon}|Y_1,Y_2,\cdots,Y_T)\approx \frac{1}{L}\sum^L_{l=1}(\sigma^{2}_{\varepsilon})^l$$
$$\hat{\sigma}^2_u=E(\sigma^2_u|Y_1,Y_2,\cdots,Y_T)\approx \frac{1}{L}\sum^L_{l=1}(\sigma^{2}_u)^l$$
where the superscript $l$ indicates the $l$th draw from the posterior distribution. It is also assumed that an appropriate burn-in period for the chain has been accounted for.

\section{Prediction\label{s4}}
\par Predicting future relations is an important and interesting problem for dynamic data, such as prediction of world trade flows, prediction of social relationship.
\par As is well known, the forecast $g(Y_1,Y_2,\cdots,Y_T)$ that minimizes the conditional mean squared error $E[(Y_{T+1}-g(Y_1,\cdots,Y_T)^2|Y_1,\cdots,Y_T]$ is the conditional expectation, that is,
$$E(Y_{T+1}|Y_1,\cdots,Y_T)=argmin_g E\left[(Y_{T+1}-g(Y_1,\cdots,Y_T)^2|Y_1,\cdots,Y_T\right]$$
Therefore, we use this function to do prediction.
\par For NVARD model, whose reduced form is (\ref{for3}), the prediction is
\begin{eqnarray}
E(Y_{T+1}|Y_1,\cdots,Y_T)&=&E(Z_{T+1}\bm{\beta}+\bm{\varepsilon}_{T+1}|Y_1,\cdots,Y_T)\nonumber\\
&=&Z_{T+1}E(\bm{\beta}|Y_1,\cdots,Y_T)\nonumber\\
&=&Z_{T+1}\hat{\bm{\beta}}\label{for8}
\end{eqnarray}
\par For VCNVARD model, whose reduced form is (\ref{for5}), the prediction is
\begin{eqnarray}
E(Y_{T+1}|Y_1,\cdots,Y_T)&=&E(Z^{(1)}_{T+1}\bm{\beta}^{(1)}+Z^{(2)}_{T+1}\bm{\beta}^{(2)}_{T+1}+\bm{\varepsilon}_{T+1}|Y_1,\cdots,Y_T)\nonumber\\
&=&Z_{T+1}^{(1)}E(\bm{\beta}^{(1)}|Y_1,\cdots,Y_T)+Z_{T+1}^{(2)}E(\bm{\beta}_{T+1}^{(2)}|Y_1,\cdots,Y_T)\nonumber\\
&=&Z_{T+1}^{(1)}\hat{\bm{\beta}}^{(1)}+Z_{T+1}^{(2)}E(\bm{\beta}_{T}^{(2)}+\mathbf{u}_{T+1}|Y_1,\cdots,Y_T)\nonumber\\
&=&Z_{T+1}^{(1)}\hat{\bm{\beta}}^{(1)}+Z_{T+1}^{(2)}\hat{\bm{\beta}}^{(2)}_T\label{for9}
\end{eqnarray}
\par It is can be seen that, for two models, the future predictions are both linear functions of coefficient estimations.

\section{Empirical Application\label{s5}}
\subsection{Data Discription\label{s5.1}}
Next, we illustrate our proposed method by a real example, world trade flows in the years 2001-2015. In this example, we use Gross Domestic Product (GDP) data and Distance data as our covariates (e.g. $X_{ijt}$ in formula (\ref{for2})) and bilateral world trade flows as our dyadic response variables (e.g. $y_{ijt}$ in formula (\ref{for2})). Gross domestic product (GDP) data can be obtained from the world bank at $http://databank.worldbank.org/data/home.aspx$. Distance data can be obtained from CEPII at $http://www.cepii.fr/CEPII/en/bdd\_ modele/bdd.asp$. Bilateral world flows can be downloaded from international trade center (ITC) at $http://www.trademap.org/Index.aspx$. Moreover, trade flows are recorded in USA dollars.
\par In order to eliminate the impact of missing data, we select a subset of the data and removed some countries so that all trade flows between the remaining countries are positive. The final number of countries we used is $n=41$. Another reason why we use these countries is that, these trade flows are relatively stable with time and subject to less affecting from special events. In 2008, the financial crisis took place, so we split the data into two parts 2001-2007 and 2009-2015. In addition, we take log for all observed data, as is common in other articles \upcite{Ward2013,Sewell2016}.
\par For the selection of weight, we intend to use equal weight, that is $\frac{1}{n-2}$ for each $w_{kql}^{oo},w_{kql}^{od},w_{kql}^{do},w_{kql}^{dd}$. Additionally, we set the coefficients of time-independent covariants, intercept and geographical distance varying with time in VCNVARD model.

\subsection{Results\label{5.2}}
\par When estimating VCNVARD model using Baysian MCMC method, the length of total MCMC chain is 300000, a burn-in of 180000 was used, leaving a chain of length 120000. To eliminate the correlation of the drawing samples, we extract a sample every 10 samples in the remaining samples.
\par We use data 2001-2006 and 2009-2014 as our training data to predict the trade flows of 2007 and 2015. Data 2001 and 2009 are initial data. We compare the in-sample (2001-2006,2009-2014) results and out-sample (2007,2015) results of different methods in root mean squared error (RMSE) evaluation criterion. The methods include univariate time series, mixed panel regression model, NVARD model and VCNVARD model.

\begin{table}
  \centering
  \renewcommand{\tablename}{Table}
  \caption{Predictive performance for 2009-2015}
    \begin{tabular}{cccccc}
    \toprule
    Method  & Univariate Time Series & Panel Regression & NVARD & VCNVARD \\
    \midrule
     RMSE-OUT  &  0.3449 &0.3193 &0.3178 &0.2869\\
     RMSE-IN   &  0.2203 &0.3337 &0.3300 &0.3169 \\
    \bottomrule
    \end{tabular}%
\label{table1}%
\end{table}%

\begin{table}
  \centering
  \renewcommand{\tablename}{Table}
  \caption{Predictive performance for 2001-2007}
    \begin{tabular}{cccccc}
    \toprule
    Method  & Univariate Time Series & Panel Regression & NARD & VCNARD \\
    \midrule
     RMSE-OUT  & 0.3726 & 0.3007 &0.2989 &0.2986\\
     RMSE-IN   &  0.2214 &0.2906 &0.2879 &0.2845 \\
    \bottomrule
    \end{tabular}%
    \label{table2}
\end{table}%

\par Table \ref{table1} and Table \ref{table2} show the predictive performance both in-sample and out-sample for two periods. The in-sample result of univariate time series is the best among all models, but its out-sample result is rather bad. Obviously, it is over-fitting. For other three models, NVARD and VCNVARD have good performances in both in-sample and out-sample. VCNVARD is better than any other models. This indicates that the proposed models capture the data generating process rather than just reflect over-fitting just like univariate time series method.
\par
\begin{figure}
  \renewcommand{\figurename}{Figure}
  \centering
  \includegraphics[width=0.45\textwidth]{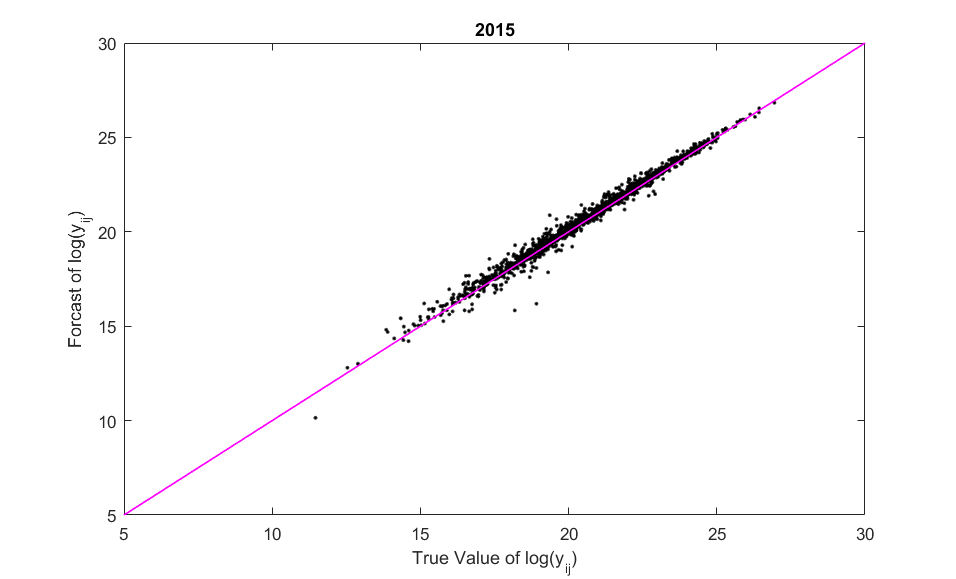}
  \includegraphics[width=0.45\textwidth]{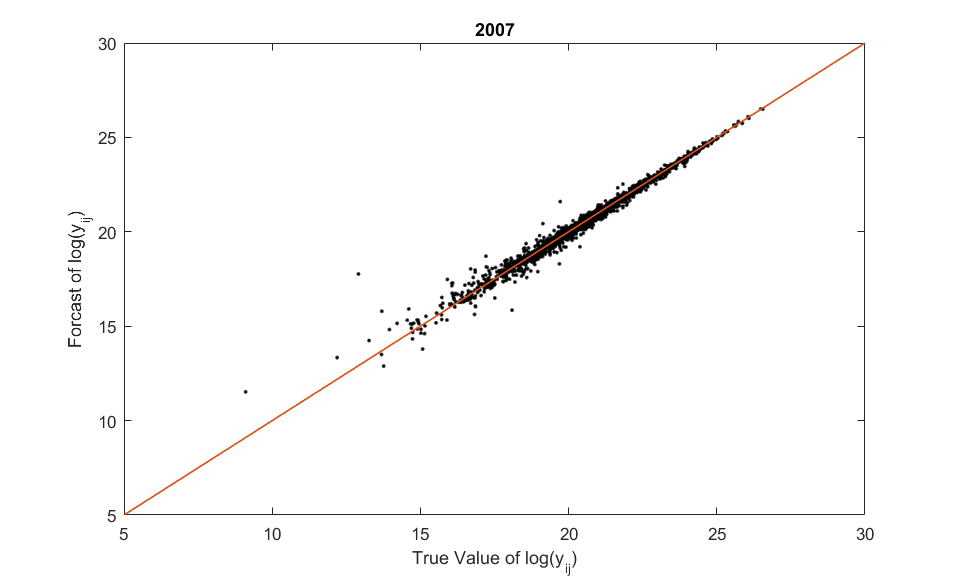}
  \caption{Comparison between True Values and  Forecast Values in 2007 and 2015}
  \label{predict}
\end{figure}

 In order to exhibit the prediction more intuitively, we make the following contrast map between true values and forecast values in Figure \ref{predict}.

It can be seen from Figure \ref{predict}, as our expectation, the predicted values are quite closed to the true values.

\section{Summary\label{s6}}
\par This paper discussed dynamic model for dyadic responses. Simple univariate time series model ignored the correlations between variables. Considering the correlation between variables, VAR model, needed to estimate too many parameters. This paper proposed NVARD model and VCNVARD model by extending NVAR model into dealing with dyadic responses. These models took both the dependencies among responses and data heterogeneous into account.
\par To conclude the article, we also discuss here several extensions of these models. Firstly, both the NVARD model and VCNVARD model require continuous responses. However, discrete responses are commonly encountered in real practice. For binary responses $\{b_{ijt}\}$ that usually appear in social network, we can use a degenerate distribution, that is,
$$b_{ijt}\sim \text{Bernoulli}(\frac{e^{E[y_{ijt}]}}{1+e^{E[y_{ijt}]}})$$
where $y_{ijt}$ is defined by (\ref{for2}) or (\ref{for4}). For counting responses $\{c_{ijt}\}$, we can use a Poisson distribution, that is
$$c_{ijt}\sim \text{Poisson}(E[y_{ijt}])$$
The above two models can be estimated by using maximum likelihood method and Bayesian method. Secondly, here we only consider the case of a small T, for a large T, the coefficients can be individual-varying, that is,
$$y_{ijt}=Z^{(1)}_{ijt}\bm{\beta}^{(1)}+Z^{(2)}_{ijt}\bm{\beta}^{(2)}_{ij}+\varepsilon_{ijt}$$
the estimate method is similar.
In addition, for the description of the dependencies, we simply use the exogenous weight matrix, that is, the dependencies between responses are linear and given in advance. In reality, how to describe the correlations of dyadic data more accurately worth further study when the data may be more complex.
\\
\\
\\
$\bf{Founding}$ Scientific Research Project of Yunnan Provincial Department of Education
(2021J0435).


%
%

\appendix
\renewcommand{\appendixname}{Appendix~\Alph{section}}
\section{\normalsize The Proof of Formula (\ref{for6}\label{A})}

Because the joint posterior distribution of $\bm{\beta},\sigma^2_{\varepsilon}$ is
$$p(\bm{\beta},\sigma^2_{\varepsilon}|Y_1,\cdots,Y_T)\propto(\sigma^2_{\varepsilon})^{-\frac{(n^2-n)T}{2}-1}\exp\left\{-\frac{\sum^T_{t=1}(Y_t-Z_t\bm{\beta})'(Y_t-Z_t\bm{\beta})}{2\sigma^2_{\varepsilon}}\right\}$$
So the marginal posterior distribution of $\bm{\beta}$ is
\begin{eqnarray*}
&&p(\bm{\beta}|Y_1,\cdots,Y_T)=\int_0^{+\infty}p(\bm{\beta},\sigma^2_{\varepsilon}|Y_1,\cdots,Y_T)d\sigma^2_{\varepsilon}\\
&\propto&\int_0^{+\infty}(\sigma^2_{\varepsilon})^{-\frac{(n^2-n)T}{2}-1}\exp\left\{-\frac{\sum^T_{t=1}(Y_t-Z_t\bm{\beta})'(Y_t-Z_t\bm{\beta})}{2\sigma^2_{\varepsilon}}\right\}d\sigma^2_{\varepsilon}\\
\end{eqnarray*}
\par Let $g=\frac{\sum^T_{t=1}(Y_t-Z_t\bm{\beta})'(Y_t-Z_t\bm{\beta})}{2\sigma^2_{\varepsilon}}$, then
\begin{eqnarray*}
&&p(\bm{\beta}|Y_1,\cdots,Y_T)\\
&\propto&\left[\sum^T_{t=1}(Y_t-Z_t\bm{\beta})'(Y_t-Z_t\bm{\beta})\right]^{-\frac{(n^2-n)T}{2}}\int_0^{+\infty}g^{\frac{(n^2-n)T}{2}-1}e^{-g}dg\\
&\propto&\left[\sum^T_{t=1}(Y_t-Z_t\bm{\beta})'(Y_t-Z_t\bm{\beta})\right]^{-\frac{(n^2-n)T}{2}}\\
&\propto&\left[\sum^T_{t=1}Y'_tY_t+\bm{\beta}'\sum^T_{t=1}Z'_tZ_t\bm{\beta}-2\bm{\beta}\sum^T_{t=1}Z'_tY_t\right]^{-\frac{(n^2-n)T}{2}}\\
&\propto&\left\{\left[\bm{\beta}-(\sum^T_{t=1}Z'_tZ_t)^{-1}\sum^T_{t=1}Z'_tY_t\right]'(\sum^T_{t=1}Z'_tZ_t)\left[\bm{\beta}-(\sum^T_{t=1}Z'_tZ_t)^{-1}\sum^T_{t=1}Z'_tY_t\right]\right.\\
&&\left.-(\sum^T_{t=1}Z'_tY_t)'(\sum^T_{t=1}Z'_tZ_t)^{-1}(\sum^T_{t=1}Z'_tY_t)+\sum^T_{t=1}Y'_tY_t\right\}^{-\frac{(n^2-n)T}{2}}\\
\end{eqnarray*}
\par Let $\mu=(\sum^T_{t=1}Z'_tZ_t)^{-1}\sum^T_{t=1}Z'_tY_t$, $v=(n^2-n)T-K$
and $$\ q=\sum^T_{t=1}Y'_tY_t-(\sum^T_{t=1}Z'_tY_t)'(\sum^T_{t=1}Z'_tZ_t)^{-1}(\sum^T_{t=1}Z'_tY_t)$$
then,
\begin{eqnarray*}
&&p(\bm{\beta}|Y_1,\cdots,Y_T)\propto\left[(\bm{\beta}-\mu)'(\sum^T_{t=1}Z'_tZ_t)(\bm{\beta}-\mu)+q\right]^{-\frac{(n^2-n)T}{2}}\\
&\propto&\left[1+\frac{1}{v}(\bm{\beta}-\mu)'\frac{v\sum^T_{t=1}Z'_tZ_t}{q}(\bm{\beta}-\mu)\right]^{-\frac{v+K}{2}}
\end{eqnarray*}
which is just the density function of multivariate t distribution with the parameters $v$, $\mu$, $\Sigma=\frac{q}{v}\left(\sum^T_{t=1}Z'_tZ_t\right)^{-1}$.

\section{\normalsize The Proof of Formula (\ref{for7})\label{B}}
From the proof of Appendix \ref{A1}, we can conclude that
\begin{eqnarray*}
&&p(\sigma^2_{\varepsilon}|Y_1,\cdots,Y_T)=\int p(\bm{\beta},\sigma^2_{\varepsilon}|Y_1,\cdots,Y_T)d\bm{\beta}\\
&\propto&\int(\sigma^2_{\varepsilon})^{-\frac{(n^2-n)T}{2}-1}\exp\left\{-\frac{\sum^T_{t=1}(Y_t-Z_t\bm{\beta})'(Y_t-Z_t\bm{\beta})}{2\sigma^2_{\varepsilon}}\right\}d\bm{\beta}\\
&\propto&(\sigma^2_{\varepsilon})^{-\frac{(n^2-n)T}{2}-1}\int \exp\left\{-\frac{(\bm{\beta}-\mu)'(\sum^T_{t=1}Z'_tZ_t)(\bm{\beta}-\mu)+q}{2\sigma^2_{\varepsilon}}\right\}d\bm{\beta}
\end{eqnarray*}
\par Let $Q=\sigma^2_{\varepsilon}(\sum^T_{t=1}Z'_tZ_t)^{-1}$, then,
\begin{eqnarray*}
&&p(\sigma^2_{\varepsilon}|Y_1,\cdots,Y_T)\\
&\propto&\exp\left\{-\frac{q}{2\sigma^2_{\varepsilon}}\right\}(\sigma^2_{\varepsilon})^{-\frac{(n^2-n)T}{2}-1}|2\pi Q|^{\frac{1}{2}}\int |2\pi Q|^{-\frac{1}{2}} \exp\left\{-\frac{1}{2}(\bm{\beta}-\mu)'Q^{-1}(\bm{\beta}-\mu)\right\}d\bm{\beta}\\
&\propto&\exp\left\{-\frac{q}{2\sigma^2_{\varepsilon}}\right\}(\sigma^2_{\varepsilon})^{-\frac{(n^2-n)T}{2}-1}|2\pi Q|^{\frac{1}{2}}\\
&\propto&\exp\left\{-\frac{q}{2\sigma^2_{\varepsilon}}\right\}(\sigma^2_{\varepsilon})^{-\frac{(n^2-n)T}{2}-1}|\sigma^2_{\varepsilon}|^{\frac{K}{2}}\\
&\propto&\exp\left\{-\frac{q}{2\sigma^2_{\varepsilon}}\right\}(\sigma^2_{\varepsilon})^{-\frac{(n^2-n)T-K}{2}-1}
\end{eqnarray*}
which is just the density function of inverse gamma distribution with the parameters $a$ and $b$, whose values are
$$a=\frac{(n^2-n)T-K}{2}$$
$$b=\frac{q}{2}=\frac{\sum^T_{t=1}Y'_tY_t-(\sum^T_{t=1}Z'_tY_t)'(\sum^T_{t=1}Z'_tZ_t)^{-1}(\sum^T_{t=1}Z'_tY_t)}{2}$$

\section{\normalsize The Full Conditional Distribution of VCNVAR Model\label{C}}
Because the joint posterior distribution of parameters and state variables is
\begin{eqnarray*}
&&p(\sigma^2_{\varepsilon},\sigma^2_u,\bm{\beta}^{(1)},\bm{\beta}^{(2)}_1,\cdots,\bm{\beta}^{(2)}_T|Y_1,\cdots,Y_T)\\
&\propto&\frac{1}{\sigma^2_{\epsilon}}\cdot\frac{1}{\sigma^2_u}\prod^T_{t=2}(\sigma^2_u)^{-\frac{m}{2}}\exp\left\{-\frac{(\bm{\beta}^{(2)}_t-\bm{\beta}^{(2)}_{t-1})'(\bm{\beta}^{(2)}_t-\bm{\beta}^{(2)}_{t-1})}{2\sigma^2_u}\right\}\\
&&\prod^T_{t=1}\prod_{i\neq j}(\sigma^2_{\varepsilon})^{-\frac{1}{2}}\exp\left\{-\frac{\left(y_{ijt}-Z^{(1)}_{ijt}\bm{\beta}^{(1)}-Z^{(2)}_{ijt}\bm{\beta}^{(2)}_t\right)^2}{2\sigma^2_{\varepsilon}}\right\}
\end{eqnarray*}
So the full conditional distributions are
\begin{eqnarray*}
&&p\left(\sigma^2_{\varepsilon}|\sigma^2_u,\bm{\beta}^{(1)},\bm{\beta}^{(2)}_1,\cdots,\bm{\beta}^{(2)}_T,Y_1,\cdots,Y_T\right)\\
&\propto&(\sigma^2_{\varepsilon})^{-1}\prod^T_{t=1}\prod_{i\neq j}(\sigma^2_{\varepsilon})^{-\frac{1}{2}}\exp\left\{-\frac{\left(y_{ijt}-Z^{(1)}_{ijt}\bm{\beta}^{(1)}-Z^{(1)}_{ijt}\bm{\beta}^{(2)}_t\right)^2}{2\sigma^2_{\varepsilon}}\right\}\\
&\propto&(\sigma^2_{\varepsilon})^{-\frac{(n^2-n)T}{2}-1}\exp\left\{-\frac{\sum^T_{t=1}\sum_{i\neq j}\left(y_{ijt}-Z^{(1)}_{ijt}\bm{\beta}^{(1)}-Z^{(1)}_{ijt}\bm{\beta}^{(2)}_t\right)^2}{2\sigma^2_{\varepsilon}}\right\}\\
&\sim&IG\left(\frac{(n^2-n)T}{2},\frac{1}{2}\sum^T_{t=1}\sum_{i\neq j}\left(y_{ijt}-Z^{(1)}_{ijt}\bm{\beta}^{(1)}-Z^{(1)}_{ijt}\bm{\beta}^{(2)}_t\right)^2\right)
\end{eqnarray*}
\begin{eqnarray*}
&&p\left(\sigma^2_u|\sigma^2_{\varepsilon},\bm{\beta}^{(1)},\bm{\beta}^{(2)}_1,\cdots,\bm{\beta}^{(2)}_T,Y_1,\cdots,Y_T\right)\\
&\propto&(\sigma^2_u)^{-1}\prod^T_{t=2}(\sigma^2_u)^{-\frac{m}{2}}\exp\left\{-\frac{(\bm{\beta}^{(2)}_t-\bm{\beta}^{(2)}_{t-1})'(\bm{\beta}^{(2)}_t-\bm{\beta}^{(2)}_{t-1})}{2\sigma^2_u}\right\}\\
&\propto&(\sigma^2_u)^{-\frac{m(T-1)}{2}-1}\exp\left\{-\frac{\sum^T_{t=2}(\bm{\beta}^{(2)}_t-\bm{\beta}^{(2)}_{t-1})'(\bm{\beta}^{(2)}_t-\bm{\beta}^{(2)}_{t-1})}{2\sigma^2_u}\right\}\\
&\sim& IG\left(\frac{m(T-1)}{2},\frac{1}{2}\sum^T_{t=2}(\bm{\beta}^{(2)}_t-\bm{\beta}^{(2)}_{t-1})'(\bm{\beta}^{(2)}_t-\bm{\beta}^{(2)}_{t-1})\right)
\end{eqnarray*}
\begin{eqnarray*}
&&p\left(\bm{\beta}^{(1)}|\sigma^2_{\varepsilon},\sigma^2_u,\bm{\beta}^{(2)}_1,\cdots,\bm{\beta}^{(2)}_T,Y_1,\cdots,Y_T\right)\\
&\propto&\exp\left\{-\frac{\sum^T_{t=1}\sum_{i\neq j}\left(y_{ijt}-Z^{(1)}_{ijt}\bm{\beta}^{(1)}-Z^{(2)}_{ijt}\bm{\beta}^{(2)}_t\right)^2}{2\sigma^2_{\varepsilon}}\right\}\\
&\sim&N(A_0^{-1}B_0,A_0^{-1})
\end{eqnarray*}
where $A_0=\frac{1}{\sigma^2_{\varepsilon}}\sum^T_{t=1}\sum_{i\neq j}(Z^{(1)}_{ijt})'Z^{(1)}_{ijt}$, $B_0=\frac{1}{\sigma^2_{\varepsilon}}\sum^T_{t=1}\sum_{i\neq j}(Z_{ijt}^{(1)})'(y_{ijt}-Z^{(2)}_{ij1}\bm{\beta}_t^{(2)})$.
\par When $t=1$,
\begin{eqnarray*}
&&p(\bm{\beta}^{(1)}_t|\sigma^2_{\varepsilon},\sigma^2_u,\bm{\beta}^{(1)},\bm{\beta}^{(1)}_{[-t]},Y_1,\cdots,Y_T)\\
&\propto&\exp\left\{-\frac{(\bm{\beta}^{(2)}_{t+1}-\bm{\beta}^{(2)}_{t})'(\bm{\beta}^{(2)}_{t+1}-\bm{\beta}^{(2)}_{t})}{2\sigma^2_u}\right\}\prod_{i\neq j}\exp\left\{-\frac{\left(y_{ijt}-Z^{(1)}_{ijt}\bm{\beta}^{(1)}-Z^{(2)}_{ijt}\bm{\beta}^{(2)}_t\right)^2}{2\sigma^2_{\varepsilon}}\right\}\\
&\sim&N(A_t^{-1}B_t,A_t^{-1})
\end{eqnarray*}
where $A_t=\frac{1}{\sigma^2_u}I+\frac{1}{\sigma^2_{\varepsilon}}(Z_{ijt}^{(2)})'Z_{ijt}^{(2)}$, $B_t=\frac{1}{\sigma^2_u}\bm{\beta}^{(2)}_{t+1}+\frac{1}{\sigma^2_{\varepsilon}}\sum_{i\neq j}(Z_{ijt}^{(2)})'(y_{ijt}-Z_{ijt}^{(1)}\bm{\beta}^{(1)})$.

\par When $1<t<T$,
\begin{eqnarray*}
&&p(\bm{\beta}^{(1)}_t|\sigma^2_{\varepsilon},\sigma^2_u,\bm{\beta}^{(1)},\bm{\beta}^{(1)}_{[-t]},Y_1,\cdots,Y_T)\\
&\propto&\exp\left\{-\frac{(\bm{\beta}^{(2)}_{t+1}-\bm{\beta}^{(2)}_{t})'(\bm{\beta}^{(2)}_{t+1}-\bm{\beta}^{(2)}_{t})}{2\sigma^2_u}\right\}
\exp\left\{-\frac{(\bm{\beta}^{(2)}_{t}-\bm{\beta}^{(2)}_{t-1})'(\bm{\beta}^{(2)}_{t}-\bm{\beta}^{(2)}_{t-1})}{2\sigma^2_u}\right\}\\
&&\prod_{i\neq j}\exp\left\{-\frac{\left(y_{ijt}-Z^{(1)}_{ijt}\bm{\beta}^{(1)}-Z^{(2)}_{ijt}\bm{\beta}^{(2)}_t\right)^2}{2\sigma^2_{\varepsilon}}\right\}\\
&\sim&N(A_t^{-1}B_t,A_t^{-1})
\end{eqnarray*}
where $A_t=\frac{2}{\sigma^2_u}I+\frac{1}{\sigma^2_{\varepsilon}}(Z_{ijt}^{(2)})'Z_{ijt}^{(2)}$, $B_t=\frac{1}{\sigma^2_u}\bm{\beta}^{(2)}_{t+1}+\frac{1}{\sigma^2_u}\bm{\beta}^{(2)}_{t-1}+\frac{1}{\sigma^2_{\varepsilon}}\sum_{i\neq j}(Z_{ijt}^{(2)})'(y_{ijt}-Z_{ijt}^{(1)}\bm{\beta}^{(1)})$.

\par When $t=T$,
\begin{eqnarray*}
&&p(\bm{\beta}^{(1)}_t|\sigma^2_{\varepsilon},\sigma^2_u,\bm{\beta}^{(1)},\bm{\beta}^{(1)}_{[-t]},Y_1,\cdots,Y_T)\\
&\propto&\exp\left\{-\frac{(\bm{\beta}^{(2)}_{t}-\bm{\beta}^{(2)}_{t-1})'(\bm{\beta}^{(2)}_{t}-\bm{\beta}^{(2)}_{t-1})}{2\sigma^2_u}\right\}\prod_{i\neq j}\exp\left\{-\frac{\left(y_{ijt}-Z^{(1)}_{ijt}\bm{\beta}^{(1)}-Z^{(2)}_{ijt}\bm{\beta}^{(2)}_t\right)^2}{2\sigma^2_{\varepsilon}}\right\}\\
&\sim&N(A_t^{-1}B_t,A_t^{-1})
\end{eqnarray*}
where $A_t=\frac{1}{\sigma^2_u}I+\frac{1}{\sigma^2_{\varepsilon}}(Z_{ijt}^{(2)})'Z_{ijt}^{(2)}$, $B_t=\frac{1}{\sigma^2_u}\bm{\beta}^{(2)}_{t-1}+\frac{1}{\sigma^2_{\varepsilon}}\sum_{i\neq j}(Z_{ijt}^{(2)})'(y_{ijt}-Z_{ijt}^{(1)}\bm{\beta}^{(1)})$.

\end{document}